# Dynamical cartography of Earth satellite orbits

Aaron J. Rosengren a,b, Despoina K. Skoulidou b, Kleomenis Tsiganis b, George Voyatzis b

a  Aerospace and Mechanical Engineering, University of Arizona, Tucson, AZ 85721, USA
b  Department of Physics, Aristotle University of Thessaloniki, 54124 Thessaloniki, Greece

**Abstract**
We have carried out a numerical investigation of the coupled gravitational and non-gravitational perturbations acting on Earth satellite orbits in an extensive grid, covering the whole circumterrestrial space, using an appropriately modified version of the SWIFT symplectic integrator, which is suitable for long-term (120 years) integrations of the non-averaged equations of motion. Hence, we characterize the long-term dynamics and the phase-space structure of the Earth-orbiter environment, starting from low altitudes (400 km) and going up to the GEO region and beyond. This investigation was done in the framework of the EC-funded ''ReDSHIFT'' project, with the purpose of enabling the definition of passive debris removal strategies, based on the use of physical mechanisms inherent in the complex dynamics of the problem (i.e., resonances). Accordingly, the complicated interactions among resonances, generated by different perturbing forces (i.e., lunisolar gravity, solar radiation pressure, tesseral harmonics in the geopotential) are accurately depicted in our results, where we can identify the regions of phase space where the motion is regular and long-term stable and regions for which eccentricity growth and even instability due to chaotic behavior can emerge. The results are presented in an ''atlas'' of dynamical stability maps for different orbital zones, with a particular focus on the (drag-free) range of semimajor axes, where the perturbing effects of the Earth's oblateness and lunisolar gravity are of comparable order. In some regions, the overlapping of the predominant lunisolar secular and semi-secular resonances furnish a number of interesting disposal hatches at moderate to low eccentricity orbits. All computations were repeated for an increased area-to-mass ratio, simulating the case of a satellite equipped with an on-board, area-augmenting device. We find that this would generally promote the deorbiting process, particularly at the transition region between LEO and MEO. Although direct reentry from very low eccentricities is very unlikely in most cases of interest, we find that a modest ''delta-v'' (DV ) budget would be enough for satellites to be steered into a relatively short-lived resonance and achieve reentry into the Earth's atmosphere within reasonable timescales (50 years).

## 1. Introduction

The solution to the debris proliferation problem, brought on by decades of unfettered space activities (Liou and Johnson, 2006), can only be found by coupling mitigation and remediation methods with a deeper understanding of the dynamical environments in which these objects reside. Recent efforts towards this more heuristic approach have explored passive means to curtail the growth rate of the debris population, by seeking to cleverly exploit the dynamical instabilities brought on by resonant perturbations, to deliver retired Earth-orbiting satellites into the regions where atmospheric drag can start their decay. Resonances, associated with commensurabilities among the frequencies of the perturbed motion, occur in profusion within the circumterrestrial phase space (q.v., Rosengren et al., 2015, and references therein). Recent work has uncovered a network of lunisolar secular resonances permeating the medium-Earth orbits (MEOs) of the navigation satellite constellations, inducing especially strong changes on the orbital eccentricity, which could be increased to Earth-reentry values (qq.v., Daquin et al., 2016; Gkolias et al., 2016). These results have put forward the idea that similar phenomena could manifest themselves throughout all circumterrestrial space, from very low-altitude orbits up to the geostationary region and beyond.

We have carried out a numerical investigation of the long-term dynamics of satellites with the goal of defining precisely these regular (long-term stable) and irregular (unstable) structures in the whole, usable circumterrestrial domain. This work was performed in the framework of the EC-funded ''ReDSHIFT'' H2020 project (q.v., Rossi et al., 2017), which aims at defining new strategies for passive debris removal, partly based on uncovering dynamically interesting regions that harbor natural trajectories that would either lead to satellite reentry on realistic timescales or constitute stable graveyards. We emphasize here the new paradigm of self-removal and reentry of satellites through natural perturbations (passive disposal), though the same analysis can be used to suggest feasible alternatives to the standard graveyards, ensuring that such storage orbits have only very small-amplitude orbital deformations over centennial timescales. While graveyard orbits should be avoided where possible, their proper definition and stability is vital for space debris mitigation (Rosengren et al., 2017).

In this work, we used a suitably modified version of the SWIFT symplectic integration package (Levison and Duncan, 1994), which has been extensively used in asteroid and Solar System dynamics. Symplectic integrators are the method of choice in such studies of (very) long-term dynamics, as they are quite efficient in integrating the non-averaged equations of motion. At the same time, they present no secular variations in the energy, thus preserving the basic premise of a conservative dynamical system. Using an extended grid of initial conditions, we characterize the dynamical architecture of the whole circumterrestrial environment from low-earth orbit (LEO) to the geosynchronous region (GEO); the latter defined as the orbit that has a period equal to one sidereal day. The results are presented in the form of an ''atlas'' of dynamical stability maps for each orbital zone, a dynamical cartography, which provides insight into the basic mechanisms that operate and dictate the dynamical evolution of satellites on long timescales (120–200 years). Particular attention has been given to the most crowded regions occupied by the current satellite and debris populations and their orbital distributions in inclination, eccentricity, and semimajor axis. This distribution was used to establish a proper grid in orbital elements for the parametric approach taken here. As a systematic study of the entire parameter space represents a formidable task with significant computational requirements, we need a model that is able to reflect the main dynamical structures in Earth satellite orbits, but still remains tractable. We present our basic physical model, based on the results of Daquin et al. (2015, 2016), and discuss the adopted simulation capabilities.

Our complete study, carried out under ReDSHIFT ( http://redshift-h2020.eu/), proceeded along two lines: a numerical exploration of the entire orbital space on a grid of initial conditions in (a; e; i) (i.e., semimajor axis, eccentricity, inclination) for selected values of the orientation-phase angles (ω; Ω) (i.e., argument of perigee, longitude of node) and a higher-resolution exploitation of the most populated areas. Semi-analytical codes that make use of appropriately averaged equations of motion to reveal the dependence of the motion on essential parameters of the system are also θsed, e.g., for the purpose of depicting the location of resonant surfaces in (a; e; i) space. The high-resolution grids of particular regions are not presented herein, apart from a few highlights for each orbital regime, as we intended to keep focus on the big picture. Our dynamical cartography underlines the complexity of circumterrestrial dynamics and displays a number of interesting and hitherto unknown features, which we briefly touch upon here, applying tools and results from dynamical systems theory, pointing out their possible role in deorbiting and mitigation strategies. Finally, we link our cartographic work to the definition of an appropriate disposal strategy for operational orbits, identifying the cost of necessary impulsive maneuvers.

**2. The Earth's orbital environment**

*2.1. The cataloged space debris*

A snapshot of the Earth's cataloged satellites and space debris in the space of orbital elements is shown in Fig. 1. Clusters correspond to particular orbital regions of interest: low-altitude orbits, which include the densely populated Sun-synchronous band near 100 deg inclination and polar orbits on the left part of the graph; geostationary-transfer orbits (GTOs) from the various launch-site latitudes (e.g., the ESA station at Kourou, KSC at Cape Canaveral, and Baikonur Cosmodrome); the critically inclined Molniya orbits in MEO; the increasingly populated Global Navigation Satellite System (GNSS) orbits; and the predominantly equatorial geosynchronous region (GEO) and supersynchronous graveyards. LEO constitutes the most densely populated orbital environment, having large concentrations in orbital inclinations in the 60deg and 110deg range, because of their use in remote sensing/weather and Earth observation missions. It is the LEO region above 600 km of altitude, which includes the crucial and crowded Sun-synchronous orbits, where special attention is also needed. At such altitudes, the natural sink mechanism of atmospheric drag is not effective on timescales dictated by official regulations (i.e., the IADC 25-year rule; Alessi et al., 2018b). Note the two large density peaks at altitudes of 800 and 1400 km.

The MEO environment, while probably not meriting the special-status treatment accorded to LEO and GEO, is host to major space-based navigation and communications infrastructure, being mainly populated by over one hundred navigation satellites in highly-inclined, near-circular orbits, the Molniya family of critically inclined, eccentric orbits, and hundreds of spent rocket bodies on GTOs. The vastness of the MEO space necessarily implies a lower spatial density and corresponding collision probabilities, compared to LEO and GEO.

The spatial density of the GEO population is several orders of magnitude below that of the LEO population, and the resulting collision probabilities are more than an order of magnitude lower. The geostationary ring, however, is the least forgiving region to space debris, because there is no natural cleansing mechanism, akin to atmospheric drag, to limit the lifetimes of debris at this altitude.

For the investigation of the Earth's magnetosphere and the interplanetary space outside of it, satellites with orbits of large semimajor axis and large eccentricity are often used (Ludwig, 1963; McComas et al., 1963). While end-of-life (EOL) disposal options are well established for missions in LEO (atmospheric decay) and GEO (near circular graveyards), existing mitigation guidelines do not fully regulate the whole, usable circumterrestrial orbital space, such as high-eccentricity science missions (HEO); e.g., NASA's Magnetospheric Multiscale Mission (MMS) (Williams, 2012) and ESA's INTErnational Gamma-Ray Astrophysics Laboratory (INTEGRAL) (Eismont et al., 2003). The non-negligible collision risks posed by these LEO-GEO transiting spacecraft has motivated both theoretical study and practical implementation (Armellin et al., 2015; Colombo et al., 2015; Merz et al., 2015). Even for the, seemingly more simple, inclined, nearly circular orbits of the navigation satellites, no official guidelines exist, and recent analyses have shown that the problem is far too complex to allow of an adoption of the basic geosynchronous graveyard strategy (Rosengren et al., 2015; Daquin et al., 2016; Celletti et al., 2016; Gkolias et al., 2016; Rosengren et al., 2017; Skoulidou et al., 2017)

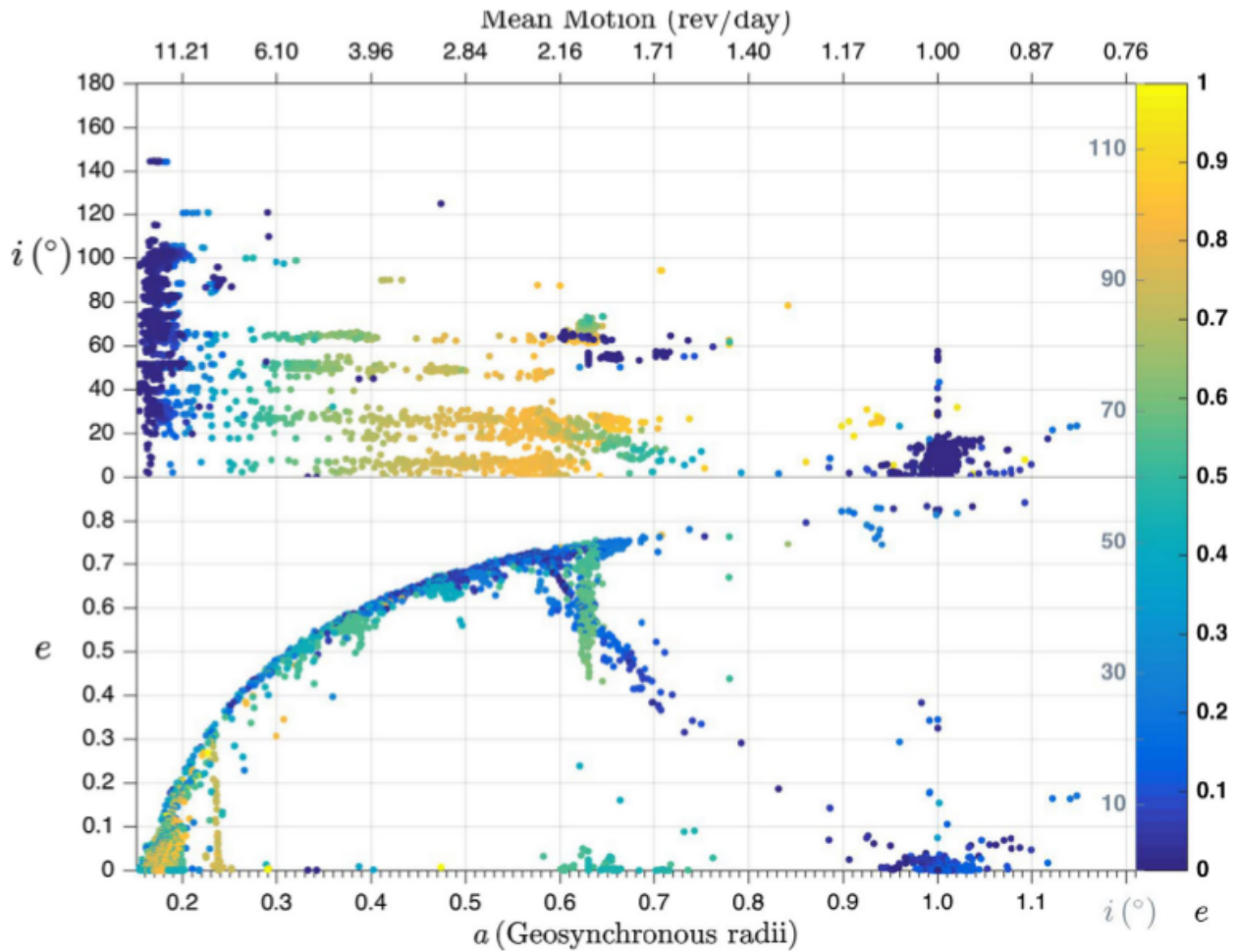

**Fig. 1.** The cataloged resident space objects in the semimajor axis–eccentricity and semimajor axis–inclination space, where the color-bar corresponds to the missing 'action' element in each two-dimensional plot (dark blue corresponds to low eccentricities or inclinations, while yellow indicates high values). ('Norad' Resident Space Object Catalog. www.space-track.org. Assessed 25 Oct. 2016). (For interpretation of the references to color in this figure legend, the reader is referred to the web version of this article.)

The precarious state of the four navigation constellations, perched on the threshold of instability, makes it understandable why past efforts to define stable graveyard orbits, especially in the case of Galileo,3 were bound to fail. A deeper understanding of the nature and consequences of resonance-overlap and chaos in these regions would have certainly helped in the early design phases of the constellations: instead of locating them in strongly perturbed dynamical environments, a modest change in their semimajor axes and/or inclinations would have placed these constellations in stable phase-space regions, likely without compromising ground visibility requirements much.

*2.2. Grid definition for numerical study*

Having an understanding of the existing space object population and their relative distributions in (a; e; i), we can properly establish the grid of initial conditions for our extensive numerical study. Such a grid naturally accounts for the relative distribution of the various Earth satellite orbits. Indeed, whereas the types of low-Earth orbits are varied, the bulk of geosynchronous are nearly circular and equatorial, excepting only the recent inclined GEO component of the Chinese BeiDou navigation system (Zhao et al., 2015). The inclinations of MEOs, on the other hand, are mainly centered around the respective launch sites or at the high values of the navigation constellations. The volume of space encompassed by these three altitude-defined regions differ substantially with MEO (altitudes between 2000 km and 35,786 km) containing roughly 34 times that of GEO (altitudes of 35,786 km ± 200 km), the latter in turn being about seven times greater than the LEO region (altitudes less than 2000 km) (Johnson, 2015).

However, we consider here a relatively coarse grid of initial conditions, spanning the complete LEO to GEO distributions in the whole range of e and i. The higher-resolution exploration, using grids specifically tailored to each orbital region based on the spatial densities of operational orbits, is presented elsewhere (Skoulidou et al., 2017; Gkolias and Colombo, 2017; Alessi et al., 2018a) or is forthcoming (Alessi et al., 2018b; Skoulidou et al., submitted for publication). Note, however, that our coarse grid also covers large parts of near-Earth space (particularly in the MEO region) that are currently not used much in operations and are typically excluded from similar studies.

Table 1 gives the basic breakdown of our cartographic survey of these regimes in terms of the nominal orbital elements covered and the initial epochs of the simulations. Note that the latter sets the initial dynamical configuration of the Earth-Moon-Sun system and orientation of the Earth, which affects the phases of the various perturbations (see, for example, Upton et al., 1959). Two epochs are used for checking (a) if the relative position of the Sun introduces important changes (i.e., semi-secular resonances; Breiter, 1999, 2001), and (b) whether the $C_{22}$ term in the geopotential introduces additional effects at LEO. The selected epochs correspond to new/full moons that are close to exact solstice, and when the Earth's longest meridian faces the Moon. Sixteen angle combinations ($\Omega$; $\omega$) were chosen to cover the most meaningful symmetries (low-order secular resonances), so that all possible high-e excursions are captured. The initial mean anomaly was set to zero, M=0, and the integration timespan was 120 years. In total, a set of roughly 19 million orbits were propagated, amounting to an equivalent of 24 years of CPU time.

**3. Problem formulation**

*3.1. Dynamical model*

It is the shape and mechanical structure of the satellite that, together with the nature and geometry of its desired operational orbit, determine which perturbing forces are dominant and which are negligible for any particular application (Montenbruck and Gill, 2000). Every distinct problem in orbital mechanics, therefore, conditions its own particular scheme of computation and requires a careful a priori examination of the dynamical situation (Daquin et al., 2016). Satellites whose altitudes fall below about 1000 km, or about 1.16 Earth radii in a (0.175 of the geosynchronous radius, $a\_GEO$ ), are significantly affected by atmospheric drag. The principal observable features of this perturbation are a secularly decreasing semimajor axis (energy), increasing mean motion, and decreasing eccentricity, so that the satellite's orbit contracts and circularizes with time (Vallado, 2013). Drag gradually wears the orbit of the satellite into the denser regions of the Earth's atmosphere, so that satellites with heights less than about 150 km remain in orbit only for a few more revolutions. A smaller, secular change in the inclination also occurs, but the perturbations in the remaining orbital elements are predominately periodic in nature (Westerman, 1963). The part of our study presented here was mainly focused on the long-term conservative effects (i.e., the "conservative back-bone" of the phase space) at altitudes larger than 1000 km. Hence, atmospheric drag was not included in our base model.

**Table 1:** Grids of initial conditions for the LEO-to-GEO study using SWIFT-SAT, for dynamical maps in a–e phase space.

| | |
|---:|:---|
| a (in aGEO ) | 0.150–1.050 |
| Da | 0.005 |
| e | 0-0.9 |
| De | 0.015 |
| i (deg) | 0-120 |
| Di (deg) | 2 |
| DΩ (deg) | {0; 90; 180; 270} |
| Dω (deg) | {0; 90; 180; 270} |
| C_R A/m (m$^2$/kg) | {0.015; 1} |

* The satellite's initial node and perigee angles were referenced with respect to the equatorial lunar values at each epoch.

The main perturbation on close satellites of the Earth (LEOs), in the absence of drag, arises from the fact that the Earth can actually be better approximated by an oblate spheroid, rather than a sphere; this asymmetry in the potential results primarily into a precession of the orbital plane about the polar axis (nodal precession) and a steady motion of the major axis in the moving orbital plane (apsidal precession). On average, for times much longer than an orbital period, (a;e;i) appear constant. Mathematically speaking, this perturbed problem is defined when one considers the gravitational potential of an oblate spheroid (the so-called "J_2 problem"), which imparts a "2nd-degree" correction on the potential of a sphere that depends only on the latitude above which the satellite moves at any time, and ignoring higher-degree gravitational harmonics due to the actual shape of the Earth. Well known applications of this particular problem are the Molniya and Tundra orbits, which are launched to an inclination of i = 63°.4; this exact value ensures that the apsidal precession freezes and, hence, the satellite's apogee can be directed above a place with the desired longitude.

The main variations caused by the third-degree zonal harmonic of the geopotential (J_3) are of long period (Vallado, 2013). Most notable are the long-term variations of e and ω, for small eccentricities, due to the appearance of a stable fixed point (the "forced eccentricity" orbit), around which all trajectories should rotate in phase space. For very small eccentricities this implies a libration of the perigee argument. Perturbations caused by higher-degree zonal harmonics are of similar nature, but of much smaller magnitude and can be ignored, except for very low-altitude orbits. As mentioned above, our main integrations focus on higher-altitude orbits and hence we have chosen to ignore these terms in the geopotential.

Unless the satellite is in a state of orbital resonance, where the satellite's mean motion is commensurable with the rotational motion of the Earth, the longitude-dependent tesseral terms of the geopotential have, generally, very little effect on the orbit and can be ignored (or, averaged) when looking on timescales much longer than one orbital period. Under resonance conditions, however, the longitudinal forces from the tesseral harmonics induce long-term changes in the semimajor axis and mean motion, leading to a libration in longitude (Ely and Howell, 1997). Contrary to higher-order zonal harmonics that essentially decay at high altitudes, tesseral resonances may become important, as the orbital period approaches commensurability with the Earth's rotation. The most important effects are associated with the lowest-order resonances and, hence, are mainly affecting the geosynchronous and GNSS regimes (Celletti and Gales, 2014). Their principal effects can be appropriately captured by a second degree and order model of the geopotential, which we include in our model.

The perturbing gravitational forces of the Moon and the Sun acting on Earth satellites cause both secular and fast periodic variations to the orbital elements (Upton et al., 1959; Hughes, 1980). For near-Earth satellites the effects of these distant perturbing bodies are often negligible,comparison to that of the Earth's oblateness, but for satellite orbits that are very elongated or have semimajor axes of several Earth radii, these third-body gravitational perturbations can change the elements of the orbit to a measurable extent, albeit on longer timescales than the Earth's oblateness. Indeed, recent works suggest that the coupled resonant effects of the Earth's oblateness and lunisolar perturbations are the most important mechanisms for delivering high-altitude, Earth-orbiting satellites into regions where atmospheric drag can give lead to orbital decay (Upton et al., 1959; Breiter, 2001; Daquin et al., 2016; Gkolias et al., 2016). Hence, these perturbations are fully taken into account in our model (in an N-body framework; i.e., without recourse to truncated Legendre expansions is done in averaged formulations).

Unlike gravitational perturbations, solar radiation pressure (hereafter, SRP) depends importantly on the size, shape and orientation of the satellite, with respect to the satellite-Sun line. The force exerted by SRP is proportional to the solar flux and to the effective area-to-mass ratio ($C_R A/m$) of the satellite, its area (A) being projected on plane perpendicular to the direction of the flux (Vallado, 2013). The direction of the acceleration is not, in general, parallel to the Sun's rays, but depends on the shape the object, its scattering law, and its orientation with respect to the Earth-Sun line. For a spherical satellite of mass m, or a flat plate object that maintains a fixed attitude perpendicular to the Earth-Sun line, the direction of the acceleration is parallel to the impinging solar rays and its magnitude is inversely proportional to the square of the distance from the Sun (McInnes, 1999); this is the so-called cannonball model, which we have adopted in the following. Note that in this model, every gravitational solar resonance is accompanied by a SRP-induced resonance, given that the mean motion of the Sun is also the frequency of SRP variability.

Other perturbations (e.g., Poynting-Robertson drag, Lorentz forces, Earth-albedo radiation pressure, etc.), with much smaller effects, are ignored in our study. The mathematical development of all perturbing effects described above is rather well known, as these were among the earliest problems to be tackled in astrodynamics (Upton et al., 1959; Musen, 1960; Brouwer and Hori, 1961; Blitzer et al., 1962; Westerman, 1963). Modern treatments can be found in any standard textbook (e.g., Montenbruck and Gill, 2000; Vallado, 2013). For this reason, and because we have not carried out any new analytical modeling, we decided to omit the mathematical details from this section.

*3.2. Simulation method*

Our "SWIFT-SAT" routine is based on the well-known mixed-variable, symplectic (MVS) integrator of Wisdom and Holman (1991), as implemented in the SWIFT numerical integration package of Levison and Duncan (1994). SWIFT has been used extensively in Solar System dynamics, and is well suited for dynamical studies of bodies with negligible mass, affected by a massive central body and other perturbing bodies, with smaller masses. The particular integration scheme is a 2nd-order, mixed-variable, symplectic routine, written in Cartesian coordinates and momenta. It is substantially faster than conventional N-body algorithms, and thus allows for an efficient, long-term integration of the precise, non-averaged equations of motion in a planetocentric (or, heliocentric) frame, without introducing artificial drift in orbital energy (secular error). Though originally designed for modeling heliocentric motions, SWIFT has been adapted in the past for dynamical studies of natural satellite systems (Nesvorny et al., 2003; Dobrovolskis et al., 2007; Morbidelli et al., 2012). This requires the Sun to be treated as a massive, distant satellite of the central planet, in order to account for its gravitational perturbations, and a rotation and translation of the coordinate system to align the fundamental plane with the planet's equator and set its origin to the center of the planet. The direct incorporation of the Sun as 'distant perturber' would significantly degrade the performance of the code, as the integration scheme is based on the assumption that the perturbers are less massive than the central object. This may be overcome by representing the motion of the Sun by an accurate ephemeris model instead. We have thus adopted the strategy of incorporating all "third bodies" in a similar fashion (here, only the Sun and the Moon), with their ephemerides taken from an accurate numerical integration of the entire solar system, produced using SWIFT for reasons of model consistency and to retain some degree of independence.

In addition, we have modified SWIFT to properly account for the dynamics in the circumterrestrial environment by incorporating the Earth's ellipticity perturbations ($J_{22}$) and solar radiation pressure (cannonball model, without shadowing effects). Note that SWIFT can be further augmented to incorporate weakly dissipative effects – of course, the symplecticity would be lost; however, slow-enough perturbations, can be assumed as adiabatic, and experience shows that this is a valid assumption – such as Yarkovsky thermal forces (Bottke et al., 2001) for asteroid dynamics or atmospheric drag for satellite dynamics, either by including them as a perturbation to the free Keplerian propagation or by coupling them with gravitational perturbations of velocities. However, we limit ourselves here to non-dissipative effects and use SWIFT-SAT to uncover the long-term dynamical behavior over an extended phase-space region, by integrating large sets of initial conditions (test particles). This global phase-space study aims at identifying where strong instabilities exist due to interacting gravitational and radiation pressure effects; these results have guided further studies of particular operational regions, using higher-resolution and more refined dynamical models (including drag), to be published elsewhere (Skoulidou et al., submitted for publication).

Our version of SWIFT-SAT accounts for the perturbations from the Earth's gravity field up to degree and order 2 (i.e., $J_{20}$ ; $J_{22}$ – easily extendable), precise lunisolar gravitational interactions, and direct radiation pressure effects. As described above, accurate ephemerides for the Sun and the Moon were computed independently, using SWIFT withfull Solar System model in the ecliptic Sun-centered frame, and then transformed into equatorial geocentric coordinates. These ephemerides were found to be in very good agreement with the high-precision ephemerides available by the Jet Propulsion Laboratory (DE421) on the timespan of interest (100–400 years); our initial conditions were taken from DE421 (Folkner et al., 2009). Note that since the time-step of the integrator, dt, is fixed (in most simulations, less than 1/250th of a sidereal day), the positions of the Moon and the Sun have been stored with full accuracy at every dt over the desired integration period and hence no interpolation was used. SWIFT-SAT was directed to remove a test particle from the integration if its distance from the Earth was less than the 400 km above the surface; this relatively large limiting h was selected on purpose, as we have chosen not to include atmospheric drag in these simulations. We underline that the units of length and time are normalized in SWIFT-SAT so that the geosynchronous distance is unity, $a\_GEO=1$ (42,164.17 km), and the period of Earth's rotation is set to 1 sidereal day (23 h 56 m 4:1 s). As a consequence, from Kepler's third law it follows that the Newtonian gravitational constant is unity ($G=1$). SWIFT-SAT is validated in Appendix A, published as Electronic Supplementary Material, against existing propagation formulations, and is shown to produce very accurate results.

*3.3. Tools from dynamical systems theory*

In interpreting large sets of simulated trajectories, corresponding to extended grids of initial conditions as used in this study, one is forced to use tools that better enable an efficient visual inspection of large portions of the phase space. This is usually achieved by constructing a series of ''dynamical maps'', where the values of specific dynamically important quantities are represented as functions of two elements. The series of maps is then parametrized by the value of a third element. Usually, these elements are either (a; e), parametrized by i, or (a; i), parametrized by e. When a specific location in (a; e; i)_0 is under study (e.g., a Galileo orbit), the two coordinates spanning the map can be the orientation angles ($\Omega,\omega$) labeled by the (a; e; i)_0 .

The quantities, whose values are usually displayed in a color-coded 2D map, are typically (a) the maximum value of a selected action-like variable, or (b) some variant of the maximal Lyapunov Characteristic Exponent (LCE), which measures the asymptotic mean rate of exponential divergence of nearby orbits; a fundamental property of chaotic motion. Action-like variables are orbital elements or combinations thereof that constitute the actions of an integrable approximation of the problem in question. We frequently denote such maps as sup-action maps (or variational map for the LCEs) (for detailed definitions and examples, see Morbidelli, 2002).Computing LCEs (or variants thereof) usually require the simultaneous integration of the system's variational equations, which are derived from the equations of motion by linearization. We note, however, that in previous studies on MEO (see, e.g., Daquin et al., 2016), it was found that typical values of the LCEs for satellite orbits imply century-long Lyapunov times. Practically, this means that chaos would only reveal itself, in the form of an orbital divergence (and, possibly, large-scale instability), on time-scales greatly exceeding those of 'practical' interest. We have verified this in a series of simulations and found that, apart from very specific situations, a detailed LCE map does not give much more practical information than a sup-action map, complemented by a map of dynamical life-times of trajectories, as the satellite's life is basically dictated by the underlying regular, secular dynamics; this is what will be presented in the following.

In the satellite case, the Delaunay elements (see Morbidelli, 2002) constitute an appropriate set of action-angle variables, where (l=M; g=$\omega$; h=$\Omega$) are the angles (coordinates) and (L; G; H)=($a^{1/2}$; L(1-$e^2)^{1/2}$; G cos i) are the actions (or, conjugate momenta). In the Keplerian approximation, as well as in the averaged ''J_2 problem'', (L; G; H) are constants of motion and denote the norm of the total angular momentum of the satellite (G), its z-component (H), and the angular momentum L of a circular, equatorial orbit with the same energy (or a). Given this property, one could use either the Delaunay elements or the usual Keplerian elements, as action-like variables. In particular, we choose to plot a variant of the maximum eccentricity ($e\_max$) attained by a propagated orbit to construct our sup-action maps. An obvious advantage of this choice is that the action has a simple dynamical and geometrical meaning. If a satellite's eccentricity grows to values higher than the one that would bring its perigee close to the upper atmosphere, then the orbit is doomed to decay.

On the other hand, each value of a corresponds to a different values of critical, reentry eccentricity $e_c$. This is why Gkolias et al. (2016) defined a different eccentricity indicator, namely, $De=(e_{max}-e_0)/(e_c-e_0)$, for each orbit with initial $e_0 < e_c$. This quantity has the property of varying between 0 and 1 (in the region $e < e_c$) and is a measure of the variation offered by the dynamics, relative to the one needed for atmospheric reentry. In practice, although $e_{max}$ is a more ''generic'' dynamical quantity, De offers sharper contrast between reentry and long-term stable solutions and highlights the finer structure of a dynamical map, related to the resonant web. Hence, we also use De in our cartography presented below (see also Appendices C and D, published as Electronic Supplementary Material).

Strictly speaking, the choice of e (or De) is not appropriate in the vicinity of a secular, lunisolar resonance, as the angular frequencies would have to obey the resonance relation and hence the corresponding angles would no longer circulate freely; this implies that neither the Keplerian nor the Delaunay elements would be appropriate actions for such a problem. However, the alternative would be to adopt a separate set of action-angle variables for every resonance in question, which would certainly limit our ability to have a ''continuous'' representation (map) and hinder a simple geometrical interpretation of the results.

Even not adopting an appropriate set of action-angle variables for each resonance, its actual effects can be depicted on a dynamical map, by plotting its locus, i.e., the set of initial conditions for which a given resonant relation between the frequencies holds (to some degree of accuracy). This can be done either by using a simplified analytical model (e.g., assuming the main perturbations to arise from the Earth's oblateness and the lunisolar secular motion alone; qq.v., Ely and Howell, 1997; Breiter, 2001; Rosengren et al., 2015; Daquin et al., 2016; Gkolias et al., 2016), or by computing the main frequencies of any propagated orbit, using a numerical (e.g., Fourier-based) technique (Tzirti et al., 2014). In Section 4.2 we employ the first technique to map the important resonances on our dynamical maps.

**4. Cartographic study from LEO to GEO**

*4.1. Global phase-space study*

In order to develop a comprehensive understanding of the long-term behavior of circumterrestrial orbits, it is necessary to cover as wide a range in orbital element space as possible while integrating the orbits over century-long timescales. Our results are presented in the form of dynamical maps, coming from integrations of millions of test particles in the circumterrestrial space and for timespans up to 120 years; this volume of simulations is necessary for the correct implementation of the passive debris removal ideology. From this dynamical survey, the main regions of long-term orbital stability and instability are mapped out as functions of (a; e; i). Much of the dynamical structure is shown to be correlated with secular and semi-secular resonances, resulting either from gravitational perturbations only, or from their coupling with solar radiation pressure.

Our dynamical atlas, presented in terms of dynamical lifetime of orbits and De maps, spans the grid given in Table 1, where $a \in (0:150; 1:050)\, a_{GEO}$ ; $e \in (0; 0:9)$, and $i \in (0; 120)$ deg. Maps were made for the two preselected epochs (JD 2458475.2433, denoted hereafter as ''Epoch 2018'', and JD 2459021.78, hereafter ''Epoch 2020'') and for several sets of orientation phase angles, in order to highlight interesting characteristics and changes that occur in the phase-space structures when such parameters are varied. Note that the same set of maps was computed for two different values of $C_R$

A/m; the large value (1 m²/kg) was chosen to represent a satellite equipped with a large sail (within current capabilities), assumed to be deployed during mission operations and to be able to keep a fixed orientation w.r.t the Sun, thus acting as a SRP-augmenting device.

Figs. 2–5 show a subset of our results, closely corresponding to the inclinations that are of most interest to the satellite operator and debris communities. Only the maps for one epoch and one phase angle combination are shown here, while a more complete atlas (including other interesting inclination values) is given in Appendix C of the Electronic Supplementary Material. Note that the gray curve appearing in all these figures is the curve of constant perigee altitude equal to 400 km, which defines, for every value of a, a critical eccentricity at which the orbit is bound to reenter into the lower atmosphere of the Earth, in our model.

Equatorial orbits of operational satellites, which typically have a relatively compact shape, and hence, a low area-to-mass ratio, are generally stable, as depicted in Figs. 2 and 3 (see also C.4 and C.5 in Appendix C). When the effective area-to-mass ratio is augmented in our model, an instability hatch opens up along the reentry corridor of the gray critical curve and continues down to low eccentricities and semimajor axes (Figs. 4 and 5). This reentry route is due to a resonance between the SRP and oblateness perturbations (Hamilton and Krivov, 1996; Chao, 2005; Colombo et al., 2012; Lantukh et al., 2015), and its location shifts towards lower altitudes with increasing inclination, approaching altitudes of 2000 km for near-polar orbits. The overall dynamical features do not change much for low-inclination values, such as those corresponding to the Kourou and Cape Canaveral (Kennedy Sace Center, KSC) launch sites (Figs. C.6–C.9), and up to around 40deg, nor do the dynamical features within these stability maps when the epoch or phase conditions are varied (at least not on this resolution). The secular dynamics may give rise to significant eccentricity excursions, but always much less than the corresponding reentry values at each a.

For high-inclinations, the maps exhibit much more complicated and interesting structures, as a result of strong interactions between different lunisolar secular resonances (Rosengren et al., 2015; Daquin et al., 2016; Celletti et al., 2016; Gkolias et al., 2016). The classical inclination-dependent-only resonances approximately occur at the well-known critical inclinations of 46.4 ; 56.1 ; 63.4; 69.0 ; 73.2 , and 90 deg , and their respective supplementary angles for retrograde motion (Cook, 1962; Vashkovyak, 1974; Hughes, 1980; Breiter, 2001; Katz and Dong, 2011; Tremaine and Yavetz, 2014). As known in general celestial mechanics, orbital resonances can be the source of both stability and instability, depending sensitively upon the choice of initial phases and, possibly, other parameters.

At the inclination value that corresponds roughly to the latitude of the Baikonur (Kazakhstan) launch site (46 deg), the most interesting observable features in Figs. 2–5 begin to show up in the MEO region and for initially high-eccentricity orbits (e > 0.6), located close to the (gray) limiting curve. The patch of relatively short-lived orbits, present even for the low A/m case, appears sensitive to the phase parameters and slightly widens with increasing A/m (i.e., increased SRP effect; cf. Figs. C.12 and C.13). For the nominal inclination values of the Galileo, GPS (56 deg), and GLONASS (64 deg) navigation constellations a number of intriguing dynamical characteristics emerge (cf. Figs. C.14–C.19). As clearly seen when following the maps from i=54deg to i=64deg , several escape hatches open up, initially at moderate and then to low eccentricities (e < 0.1), very close to the nominal semimajor axes of these constellations. This is the result of the principal lunisolar secular resonances. Note that, as seen in the lifetime maps, reentry within 80 yr is natural for orbits almost exactly located at the Galileo constellation, but of course for higher eccentricity (e > 0.1) than the one of the satellites. Note also that the shape and orientation of these reentry hatches depend strongly on the orientation of the orbital plane and apsidal line and the chosen epoch, as can be seen in the following figures (also noted in Daquin et al., 2016).

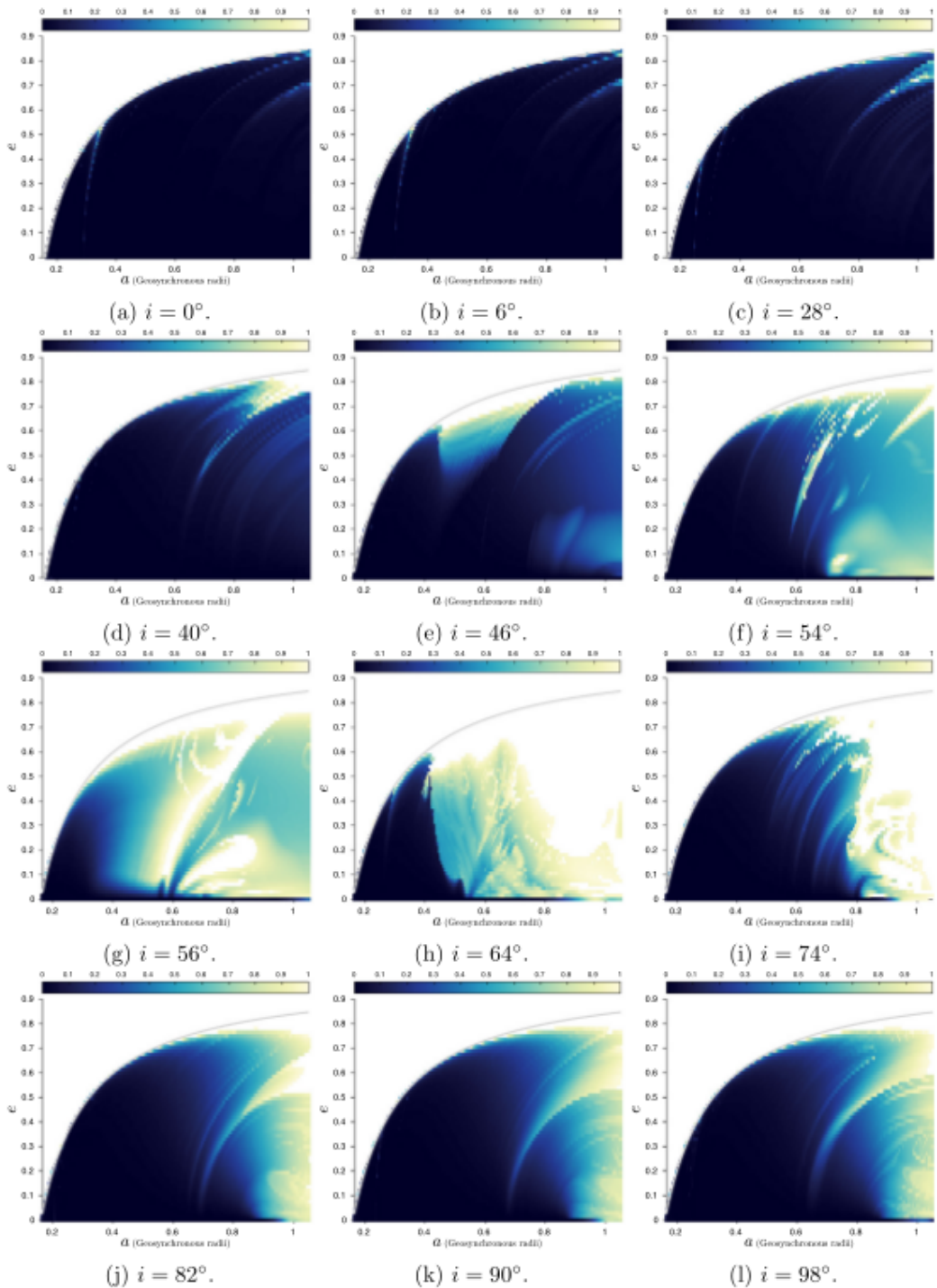

**Fig. 2.** De maps of the global LEO-to-GEO phase space in a–e as a function of i, for DΩ=270 deg ; Dω= 90 deg, epoch 2018, and C_R A/m=0.015 m²/kg. The colorbar is from 0 to 1. (For interpretation of the references to color in this figure legend, the reader is referred to the web version of this article.)

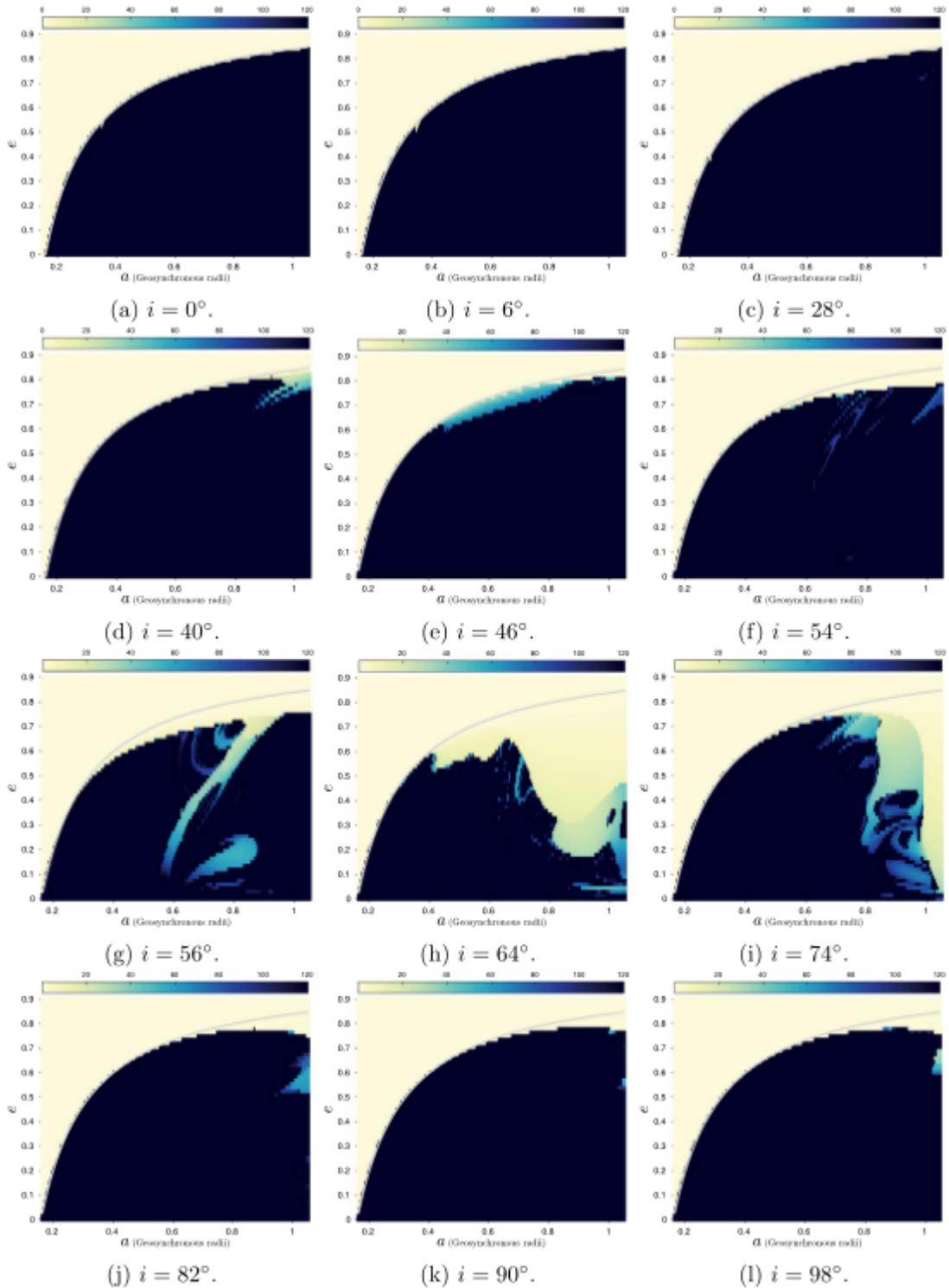

**Fig. 3.** Lifetime maps of the global LEO-to-GEO phase space in a–e as a function of i, for DΩ=270 deg; Dω=90deg , epoch 2018, and C_R A/m=0.015 m²/kg. The colorbar is from lifetime 0 to 120 years. (For interpretation of the references to color in this figure legend, the reader is referred to the web version of this article.)

The high-density MEO region of i=64deg and LEO region at i=74deg , become mostly unstable as we move away from the Earth and approach GEO altitudes, as shown also in Figs. C.20–C.23. In fact, the high-i extended GEO regionaround i ~60–80deg is mostly unstable, with dynamical lifetimes smaller than 30 years. On the other hand, the maps corresponding to polar and Sun-synchronous inclinations (Figs. C.24–C.27) do not exhibit many reentry regions, especially for the low-altitude orbits where they are mainly used, except for the case of augmented A/m (e.g., by a large sail).

As seen in these graphs, the high-LEO and equatorial GEO regions show a natural deficiency of reentry solutions, even for eccentricities slightly higher than zero. On the other hand, the 'enhanced SRP' simulations suggest that reentry solutions may be present in the vicinity of high LEOs, while the lunisolar complex provides escape routes for MEOs. Note, however, that atmospheric drag, which can be decisive for the reentry of LEO (and, likely, GTOs) for altitudes smaller than 1000 km (qv. Wang and Gurfil, 2016), is not included in our simulations. Hence, our main results for high altitudes are still valid even in the LEO zone (e.g., the escape hatch at altitudes h > 2000 km) for low to moderate eccentricities, but for the densely populated LEO region a model incorporating drag should be used. These features will be explored in detail in upcoming publications, where higher-resolution simulations for different densely-populated regions are used, in a model that incorporates drag.

*4.2. Identification and effect of principal resonances*

The most important dynamical feature of the periodically perturbed Kepler problem is the existence of resonances. When resonances occur, the eccentricity is the most important orbital element in terms of long-term stability, as any change in e affects the perigee radius, which influences the satellite's lifetime. In the satellite problem, the most important secular resonances (i.e., excluding tesseral resonances between the mean motion of the satellite and the Earth's rotation rate) are the gravitational lunisolar resonances, and, for relatively high values of A/m, the SRP-related resonances. Any resonance is defined as the locus of points in phase space for which a specific linear combination of frequencies becomes null. In our case, the principal resonances can be cast in the form

$$\dot{\psi} = j\dot{\omega} + k\dot{\Omega} + l\dot{\Omega}_M + m\, n_S \approx 0 \quad (1)$$

where ω is the critical angle of the resonance, j; k; l; m are integers, XM denotes the node of the Moon, and n_S is the apparent mean motion of the Sun. The locus of the resonance in orbital space can be found by inverting the above relation, having first chosen an analytical expression for the precession frequencies. To lowest order, the resonant phase space is topologically similar to that of a pendulum; i.e., the critical angle can either librate or circulate, if the orbit tarts inside or outside the separatrix, respectively. If ω involves both ω and Ω, then coupled, nonlinear variations of the conjugate actions (i.e., e and i) are to be expected. Finally, let us add here that in a fully nonlinear description of the problem, a resonance cannot be treated in isolation from nearby resonances, and their interaction leads to the emergence of chaotic motions around the separatrix, characterized by irregular transitions of w between libration and circulation (see Daquin et al., 2016; Celletti et al., 2016, for a more detailed analytical treatment of lunisolar resonances; also reviewed in Appendix B).

The center of each lunisolar and radiation pressure semi-secular and secular resonance (for both prograde and retrograde orbits) may be approximately defined in the action phase space, as shown in Figs. 6–8. For low A/m values (as in this study), the secular frequencies of the perigee and node can be very well approximated by a simple J_2 model for the Earth's potential. Similarly, the correction given by the effects of the Moon/Sun is very small, except for high-eccentricity orbits. Hence, for the purposes of simply superimposing the resonances loci on our dynamical maps (see Appendix C), we have used here the simple, analytical formula of the averaged J_2 problem.

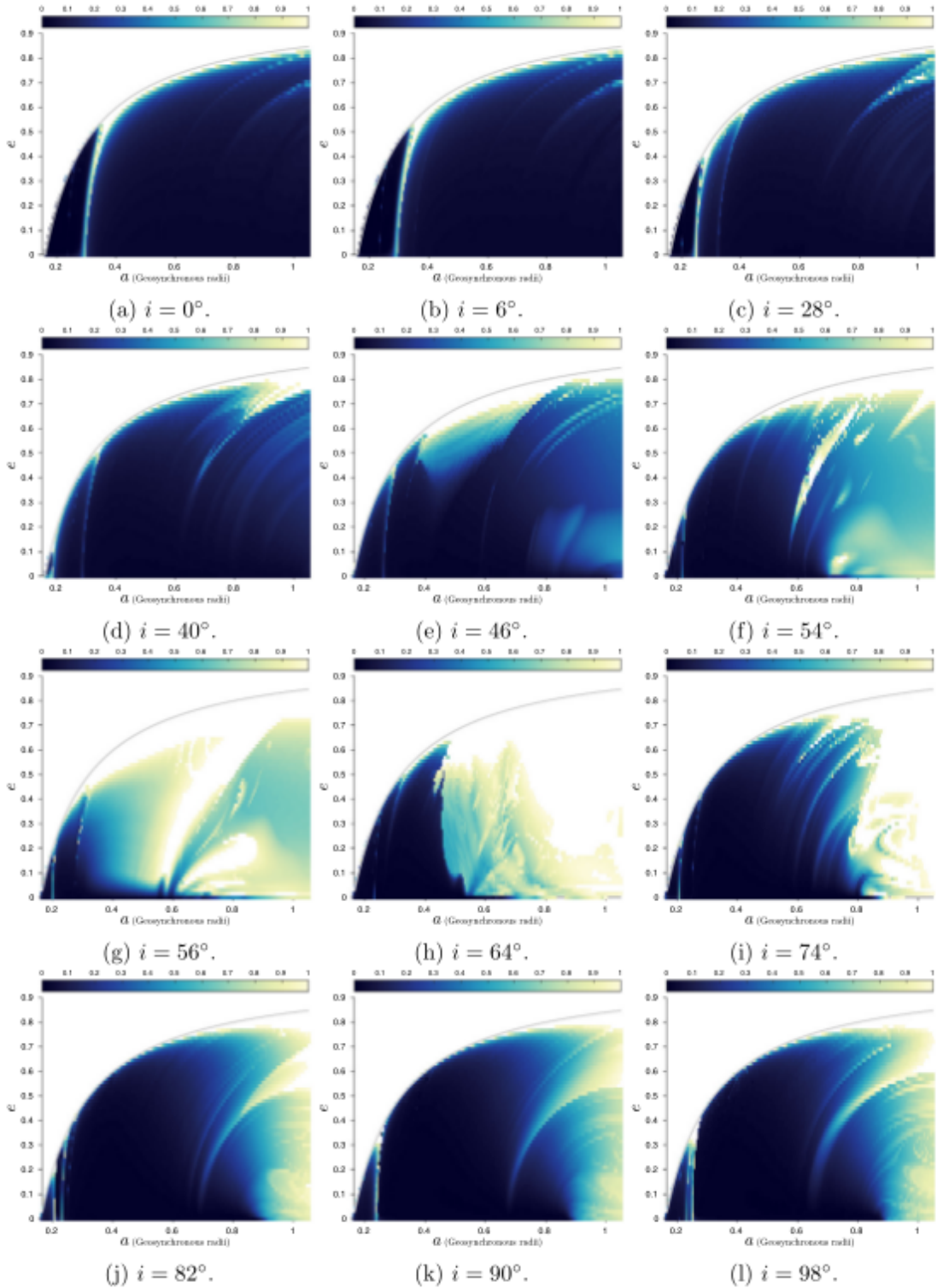

**Fig. 4.** De maps of the global LEO-to-GEO phase space in a–e as a function of i, for DΩ=270deg ; Dω=90deg , epoch 2018, and C_R A/m = 1 m2 /kg. The colorbar is from 0 to 1. (For interpretation of the references to color in this figure legend, the reader is referred to the web version of this article.)

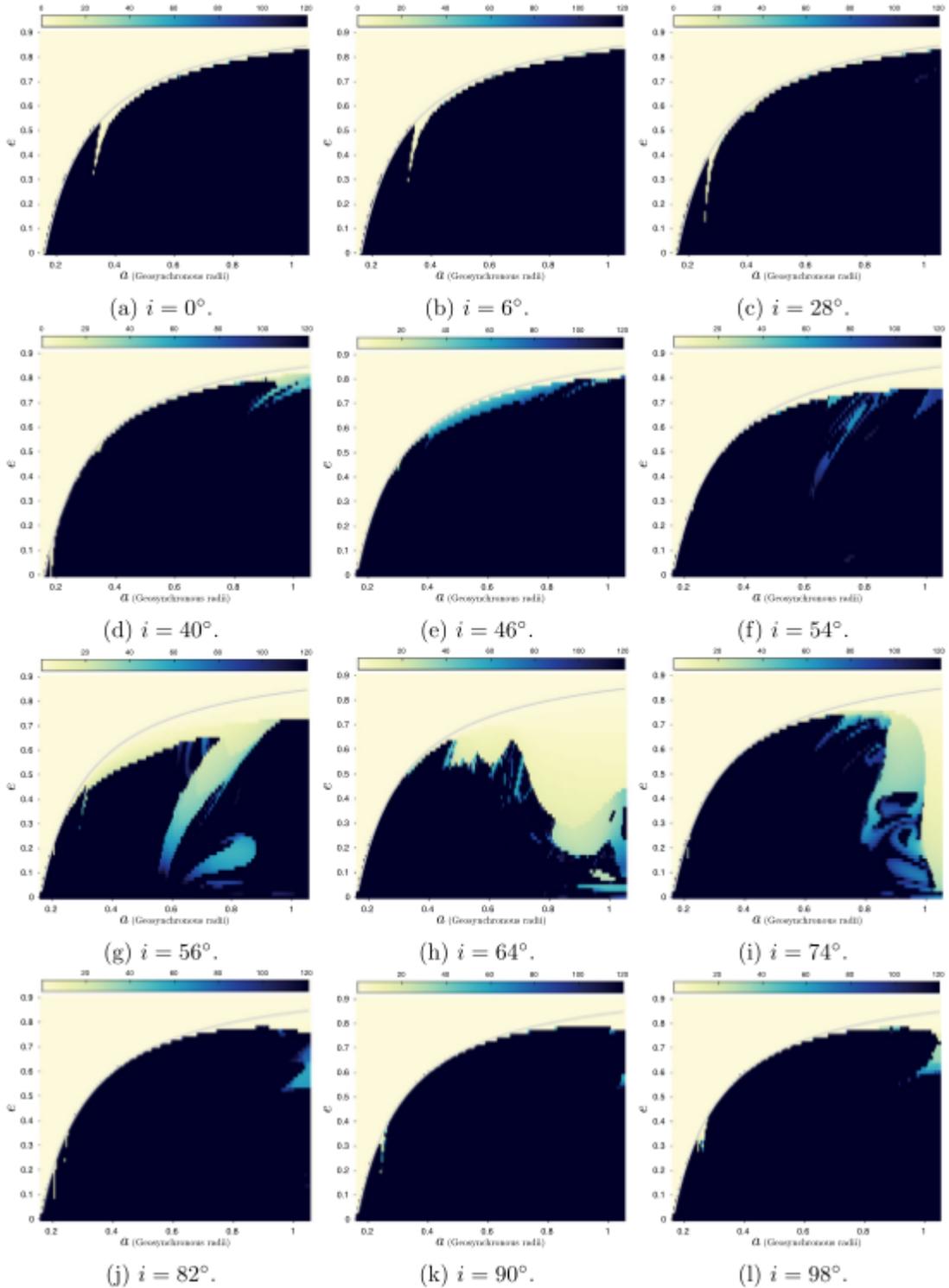

**Fig. 5.** Lifetime maps of the global LEO-to-GEO phase space in a–e as a function of i, for DΩ= 270deg ; Dω= 90deg, epoch 2018, and C_R A/m= 1 m²/kg. The colorbar is from lifetime 0 to 120 years. (For interpretation of the references to color in this figure legend, the reader is referred to the web version of this article.)

Fig. 6 maps the resonance web in the (i;a) region, for e=0 and for e=0.6. This highlights the symmetry with respect to i=90deg and the progressive ascent of the web to higher values of a, for higher eccentricities. The different colors correspond to different combinations of the integer coefficients (j; k; l; m) in Eq. (1); the color scheme follows that of Rosengren et al. (2015), to which we refer for the omitted details. Part of the resonance web lies below the surface of the Earth (gray-shaded area) and thus has no physical meaning, but is given to highlight the multiplet-like structure. One of the most striking features in this figure is the multiple crossings that occur between various resonances in the MEO region, even for nearly-circular orbits. In fact, it is this overlapping at i~56deg that explains the presence of the instability hatch at low eccentricities, at the Galileo constellation value of a. A different projection of the web is given in Fig. 7, for two different values of a at which several lunisolar resonances appear to cluster. Again, the plot for a=0.681 a_GEO suggest multiple resonance crossings down to low eccentricities, something which is much less prominent in the corresponding a=0.303 a_GEO graph. Finally, in Fig. 8, the resonant web is projected on the (a; e) plane, for equatorial (i=0) and Galileo-like (i=56deg) orbits. For equatorial orbits, secular resonances cluster near 0.35 a_GEO and span the high-a region from 0.5 to 1 a_GEO. However, as can be seen also in Fig. 6(a) these resonances are rather well-separated and this explains the smoothness of the secular dynamics even in the eccentric domain around aGEO . On the other hand, near i=56deg secular resonances come very close to each other. This is difficult to see directly in Fig. 8(b), but it is implied by the fact that fewer curves are seen in the MEO region, than in panel (a); this is because some of the curves are now almost overlapping each other. Looking back to Fig. 6, one can see that, indeed around that particular inclination, there are several crossing points of different families of resonances, from 0.5 to 0.8 a_GEO.

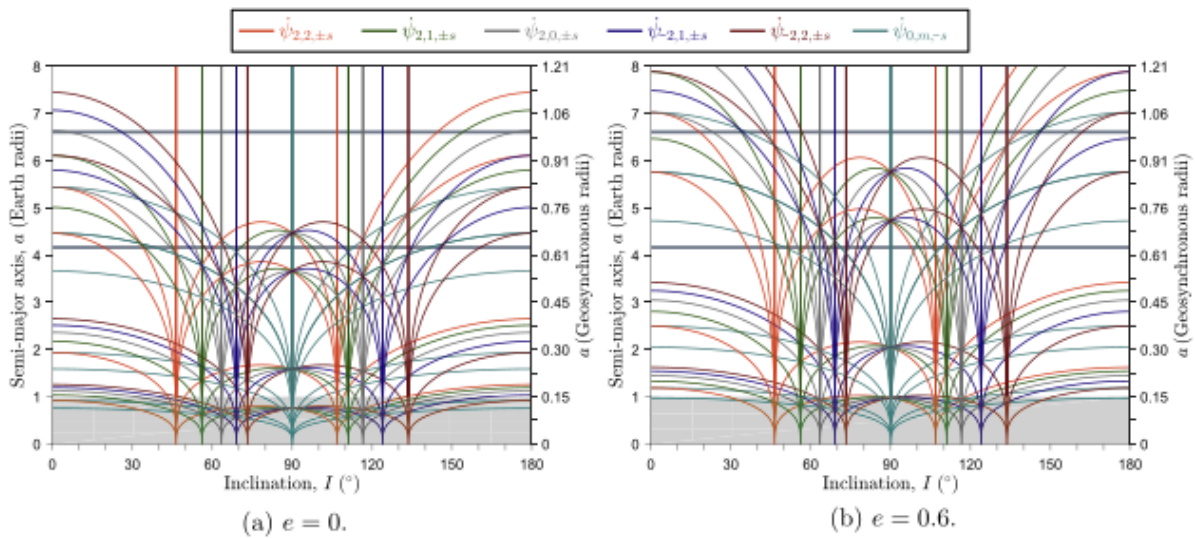

**Fig. 6.** Location of all principal lunisolar and radiation pressure resonances in the inclination–semimajor axis plane, for circular and eccentric satellite orbits. The upper horizontal lines represent the location of the 1:1 (GEO) and 2:1 (GPS) tesseral resonances, respectively. The gray area corresponding to inside the Earth is unphysical, but shown to highlight the resonant skeleton structure. The color scheme for the curves follows that of Reference (Rosengren et al., 2015) (For interpretation of the references to color in this figure legend, the reader is referred to the web version of this article.)

After mapping the resonant web and understanding its 10long-term dynamical effects, we can now seek to define innovative strategies for designing satellite demise orbits (i.e., reentry solutions) that make use of this dynamical structure. In particular, we wish to understand whether we could use those natural eccentricity-growth corridors in phase space that could promote orbit decay within realistic timescales. As seen in this section, such disposal hatches naturally lie along resonant lines or in regions where several resonance lines cross (resonance overlap). In Fig. 9, one can see an example for two particles, starting from i=56deg (left) and i=64deg (right), of how the overlapping of secular resonances affect on the evolution of eccentricity and inclination.

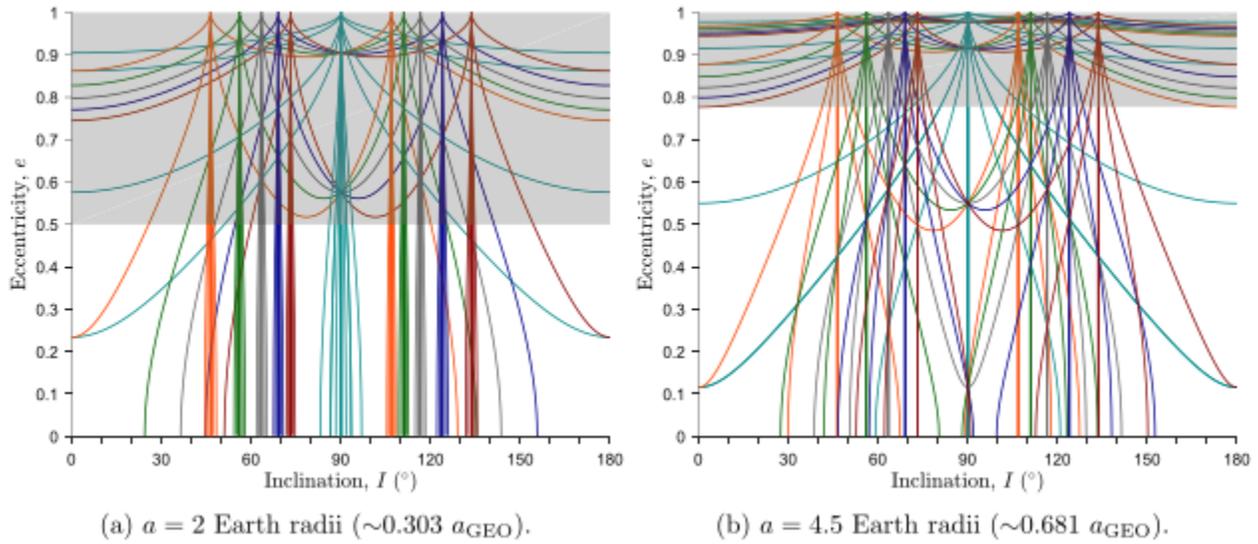

**Fig. 7**. Location of all principal lunisolar and radiation pressure resonances in the inclination–eccentricity plane, for satellite orbits. The gray area corresponding to inside the Earth is unphysical, but shown to highlight the resonant skeleton structure. The color scheme for the curves follows that of Reference (Rosengren et al., 2015), to which we refer for details. (For interpretation of the references to color in this figure legend, the reader is referred to the web version of this article.)

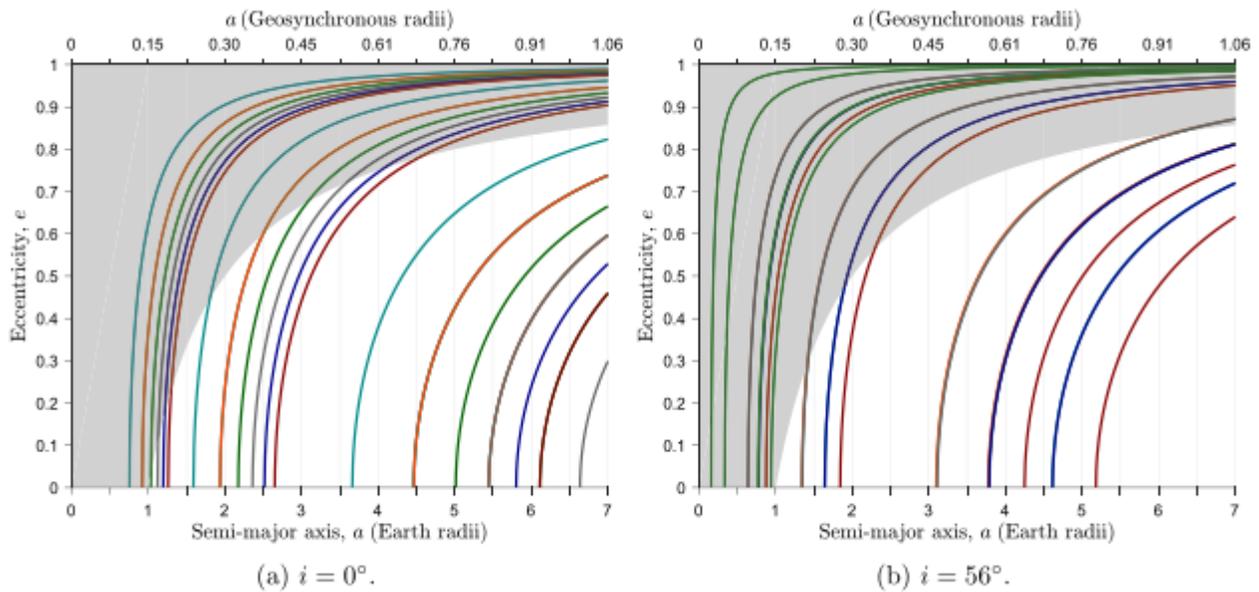

**Fig. 8.** Location of all principal lunisolar and radiation pressure resonances in the semimajor axis–eccentricity plane, for equatorial and inclined satellite orbits. The gray area corresponding to inside the Earth is unphysical. The color scheme for the curves follows that of Reference (Rosengren et al., 2015), to which we refer for details. (For interpretation of the references to color in this figure legend, the reader is referred to the web version of this article.)

# 5. Using the Dynamical Atlas for designing satellite disposal

Our results, apart from providing a graphical overview of the long-term dynamics around the Earth, also contain essential dynamical information; namely, eccentricity variations and dynamical lifetimes for all orbits integrated. The long-term eccentricity variations are essential for deciding whether an orbit that does not escape within the 120 years of the integration can constitute a viable grave-
yard solution. For escaping orbits, on the other hand, short dynamical lifetimes are desirable. Several escape routes are seen on our maps, in particular around the high-LEO/low-MEO interface and near the inclined MEO population with reentry times between 10 and 120 years; the eccentricity growth is associated to the action of lunisolar resonances. However, escape solutions appear, in general, for eccentricities larger than ~0.05 (except for the SRP-related hatch for near-polar, high-LEOs). This means that typical operational satellites (e ~= 0) would not evacuate their operational region unassisted, even after very long times. Moreover, long dynamical lifetimes, found, e.g., in the MEO region, may actually lead to a negative result, as a MEO satellite disposed on such a demise orbit on purpose may actually be repeatedly crossing the protected regions (LEO below 2000 km and the GEO ring) for long times. Hence, it is important to check whether the reentry solutions found can actually be used and how.

The definition of an appropriate disposal strategy for any operational orbit, essentially involves calculating (optimal) maneuvers needed to reach the desired disposal orbit (Gobetz and Doll, 1969). The first step in this direction was performed recently by Armellin and San-Juan (2018) in the case of Galileo; however, their multi-objective optimization approach is beyond our scope here. Alternatively, one can ask which orbits can be reached for a given fuel constraint (DV), starting from some initial operational orbit (Marec, 1979). The study of optimal orbit transfers dates back to the 1960s, where fuel-optimal solutions were found for both single velocity impulse and multiple-impulse (e.g., Hohmann-type) trajectories (Gobetz and Doll, 1969; Marec, 1979). With a given DV, it is possible to reach, in the three-dimensional space of the variations Da; De; Di, all of the orbits situated in a certain volume called the reachable domain. The optimal transfers correspond to the boundary of the reachable domain. Modern methods for determining this boundary are often quite involved (Xue et al., 2010; Holzinger et al., 2014).

An approximate calculation of the reachable domain can be performed by using the well-known Gauss equations, assuming a given initial orbit (a; e; i) and a single, impulse of fixed magnitude DV and varying orientation. Hence, we can chose an 'operational orbit' as starting point and, given a DV budget, search for reentry solutions that are contained in our maps and, at the same time, lie within the boundaries of the reachable domain. We can further restrict our search, by requiring co-planar transfers; this means looking only into maps with the same value of (i; Ω) as the operational orbit, but allow for a change in ω. Similar computations can be done, e.g., for Hohmann transfers between co-planar and co-axial ellipses that intersect (or not) the operational orbit. We can then sum up our results and represent all solutions found in a DV-lifetime graph.

The above procedure was implemented for a typical GPS, a typical Galileo and a typical GLONASS orbit, all with initial e= 0.0001 and A/m = 0:015 $m^2$/kg. The results are shown in Fig. 10. Every point on each graph is a (co-planar) reentry solution, contained in our maps, that can be reached from the respective initial orbit with a single- or two-burn (Hohmann) impulsive maneuver that has a fuel cost of DV <=3.5 km/s. Of course this DV limit is unrealistic and a reasonable DV would not exceed ~200 m/s in practice. However, we have extended here the graph to this high DV max value, simply to show the structure of the solutions' space. In each graph one can see two 'V-shaped' clouds of points; these correspond to the single- and two-burn transfers. The lower envelopes of these V-shaped clouds consist of the respective optimal solutions (known as atmospheric reentry Pareto fronts; Armellin and San-Juan, 2018).

As can be seen in all panels of Fig. 10, orbits that reenter within T_r < 25 years can be found, but correspond to DV much greater than 300 m/s (unrealistic in practice), with the exception of a few cheaper solutions found for GLONASS – note, however, that these solutions may not be reachable for every set of orientation angles. For the GPS and Galileo orbits, disposal orbits with reentry times as small as ~60 years can be reached with DV<=100 m/s. These solutions may constitute a good compromise between waiting time and fuel cost. On the other hand, for GLONASS, the cost decreases very slowly allowing for longer T_r ; this means that a relatively high cost needs to be paid, irrespective of the chosen solution. Hence, for GLONASS, looking for such demise orbits may not be a viable strategy.

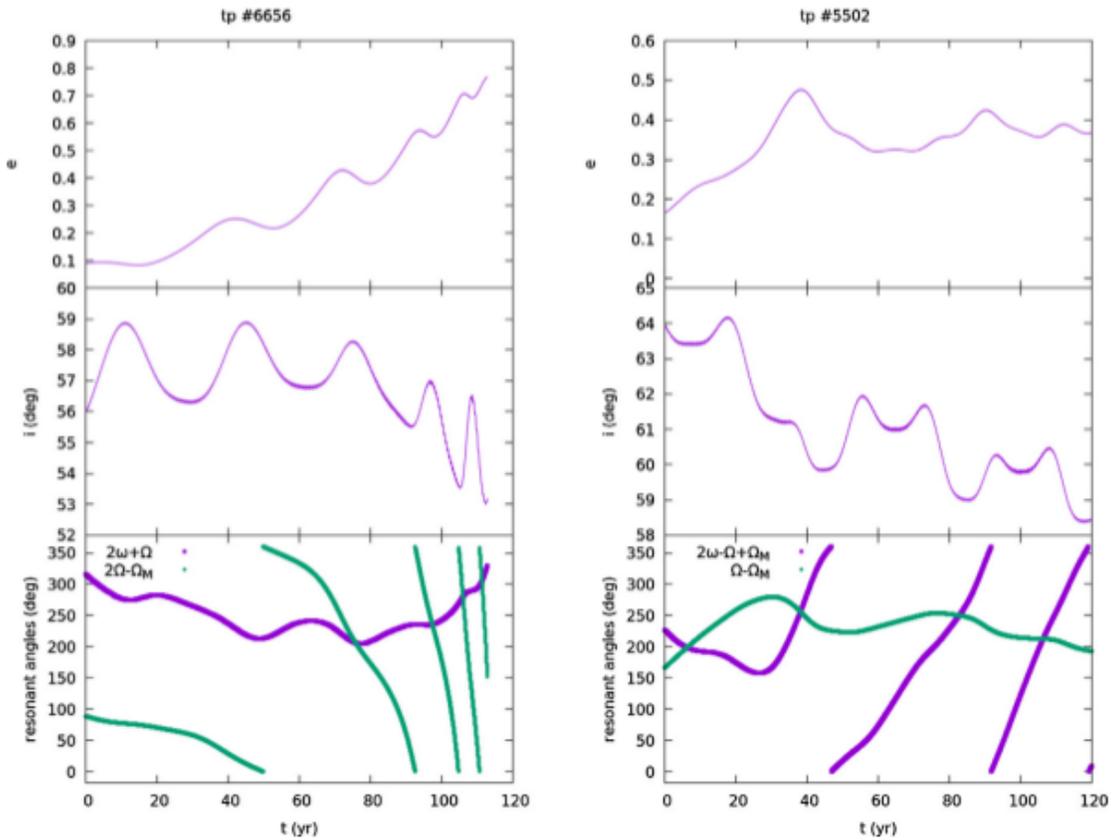

Fig. 9. Evolution of eccentricity (*top*), inclination (*middle*) and of two resonant angles (*bottom*) for two particles. The one on the *left* (labeled tp #6656) starts from $i = 56°$ and $a \sim a_{Galileo}$ (with $e = 0.090$, $\Omega = 102.83°$, $\omega = 286.50°$, $M = 0$), and crosses resonances $\dot{\psi} = 2\dot{\omega} + \dot{\Omega} \approx 0$ and $\dot{\psi} = 2\dot{\Omega} - \dot{\Omega}_M \approx 0$. The one on the *right* (tp #5502) starts from $i = 64°$ and $a \sim a_{GLONASS}$ (with $e = 0.165$, $\Omega = 282.83°$, $\omega = 196.50°$, $M = 0$), and crosses resonances $\dot{\psi} = 2\dot{\omega} - \dot{\Omega} + \dot{\Omega}_M \approx 0$ and $\dot{\psi} = \dot{\Omega} - \dot{\Omega}_M \approx 0$.

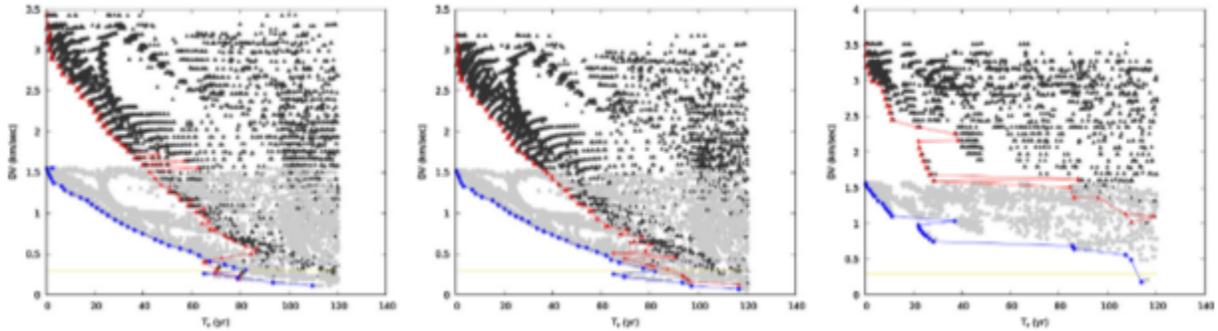

**Fig. 10.** DV -lifetime maps for reentry solutions found around typical GPS (left), Galileo (middle) and GLONASS (right) orbit. The yellow line corresponds to DV = 300 m/s.

The percentage of reentry solutions that have DV<=300 m/s is between 0.19 and 0.21 for Galileo, 0.16 and 0.18 for GPS, and 0.05 and 0.03 for GLONASS, weakly depending on the map's epoch and assumed A/m value. The reentry orbits were checked with respect to the cumulative time that they spend orbiting inside the protected LEO/GEO regions. It turns out that, even for very high $T_r$ values, the cumulative time spent in LEO/GEO is typically of order 0.5 years, with a few outliers spending ~2 years in total. Only for initially high-eccentricity (i.e., GTO-like) orbits — which can only be reached with high DV s and, therefore, would anyway be impractical — the median time spent in LEO is ~2 years and a few outliers can reach up to ~20 years. Hence, in no case do long reentry times actually violate the IADC 25-year rule.

Note that the dynamical maps presented here, while very helpful in providing a global view of the long-term dynamics of satellites around the Earth, are not of sufficiently high resolution for computing the optimal maneuver for a given operational orbit. The construction of higher-resolution maps, dedicated to the crowded LEO, MEO and GEO maps, as well as their use for designing disposal orbits are one of the major goals of the ReDSHIFT project and are to be presented in detail in upcoming publications. However, we believe to have adequately demonstrated the feasibility of this procedure, using the maps presented in this paper and applying it to typical GNSS orbits.

## 6. Conclusion

We presented the most complete to date dynamical atlas of the entire usable circumterrestrial space, characterizing the long-term dynamical behavior of Earth satellites from LEO to GEO and beyond. About 20 million orbits were propagated for 120 years, in a model that included 3rd-body, lunisolar perturbations, a 2nd-degree and order geopotential and a cannonball SRP model. The computations were repeated for two initial epochs and assuming two A/m values; one corresponding to a 'nominal' satellite and the other to a satellite equipped with an area-augmenting device, such as a sail within current technological capabilities. The non-averaged equations of motion were integrated using a well-established symplectic scheme, properly tested against other known techniques. The eccentricity variations and the dynamical lifetime of the orbit (deemed as 'reentry' if it approached the surface of the Earth within 400 km in our drag-free model) were used as the main dynamical indicators. A simple analytical model was used in order to map the web of lunisolar secular resonances in orbital elements space, which can be projected on a selected 2-D map.

The resolution of our grid of initial conditions was relatively high in e and i, but only a few selected combinations of (Ω,ω) were used. Hence, a more complete analysis is needed and is currently under way, with particular focus on the current densely populated regions. Nevertheless, as the particular angle combinations chosen account for the main symmetries of the problem, we believe that our atlas largely covers the long-term dynamics. With its limitations, our atlas gives a full account of all phase-space regions that could be useful for defining innovative disposal strategies for future missions. These include regions of low-amplitude eccentricity variations (graveyards) but also regions from which direct reentry to the lower atmosphere of the Earth is possible. We have analyzed extensively the latter, showing their relation to overlapping secular resonances.

In addition, we have demonstrated how our dynamical atlas could be used to design the disposal of a GNSS-like satellite towards an orbit that leads to direct reentry and respects a given DV constraint. In fact, we found that solutions with dynamical lifetime 60 y can be reached with a moderate DV ~ 100 m/s. Moreover, such solutions do not violate the IADC 25-year rule, as the cumulative time spent in LEO is very small. In principle, a similar method could be used for the high-LEO region.

Active debris removal, apart from the daunting obstacles in practical engineering and the difficult financial, political, and legal challenges that it represents, is widely seen by the debris community as the only suitable option to prevent the self-generating Kessler phenomenon. Such drastic measures, however, should be reassessed on account of our increasing knowledge on the long-term effects of the principal resonances in near-Earth space. It was shown here that many of the resonances found in our study can be exploited on decadal-to-centennial timescales to effectively remove satellites from crowded regions and their long lifetime orbits. A proper definition of the end-of-life strategy that takes into account the long-term dynamics, in conjunction with relatively low-DV maneuvers and possibly an A/m-augmenting device (e.g., sail), from the early design phase can possibly yield a self-correcting mechanism, much needed to sustain the space environment. In this regard, choosing the most effective strategy able to drive a given satellite towards a reentry solution must be based on a deeper understanding of the natural dynamical environment of the specific orbital region, in which the satellite resides.

**Acknowledgements**

This research is funded in part by the European Commission Horizon 2020 Framework Programme for Research and Innovation (2014–2020), under Grant Agreement 687500 (project ReDSHIFT; http://redshift-h2020. eu/). We would like to acknowledge the ReDSHIFT team for many discussions and for their internal review of this work. Special thanks goes to I. Gkolias, D. Amato, J. Daquin, and F. Gachet for many insightful and motivating conversations. Numerical results presented in this work have been produced using the Aristotle University of Thessaloniki (AUTh) Computer Infrastructure and Resources and the authors would like to acknowledge continuous support provided by the Scientific Computing Office. The work of D.K. Skoulidou was also supported by General Secretariat for Research and Technology (GSRT) and Hellenic Foundation for Research and Innovation (HFRI).

**Appendix A. Supplementary material**
Supplementary data associated with this article can be found, in the online version, at https://doi.org/10.1016/j.asr.2018.09.004.